\documentclass[twocolumn,superscriptaddress,nobalancelastpage,10pt,pra,letter]{revtex4}
\usepackage{amsmath}
\usepackage{amsthm}
\usepackage{amssymb}
\usepackage{amsfonts}
\usepackage{graphicx}
\usepackage{wasysym}
\usepackage{mathrsfs}
\usepackage{yfonts}
\usepackage{bbold}
\usepackage{verbatim}
\usepackage{subfigure}
\usepackage{color}
\usepackage{setspace}
\usepackage{multirow}
\usepackage{blkarray}
\makeatletter
\let\save@mathaccent\mathaccent
\newcommand*\if@single[3]{%
	\setbox0\hbox{${\mathaccent"0362{#1}}^H$}%
	\setbox2\hbox{${\mathaccent"0362{\kern0pt#1}}^H$}%
	\ifdim\ht0=\ht2 #3\else #2\fi
}
\newcommand*\rel@kern[1]{\kern#1\dimexpr\macc@kerna}
\newcommand*\widebar[1]{\@ifnextchar^{{\wide@bar{#1}{0}}}{\wide@bar{#1}{1}}}
\newcommand*\wide@bar[2]{\if@single{#1}{\wide@bar@{#1}{#2}{1}}{\wide@bar@{#1}{#2}{2}}}
\newcommand*\wide@bar@[3]{%
	\begingroup
	\def\mathaccent##1##2{%
		\let\mathaccent\save@mathaccent
		\if#32 \let\macc@nucleus\first@char \fi
		\setbox\z@\hbox{$\macc@style{\macc@nucleus}_{}$}%
		\setbox\tw@\hbox{$\macc@style{\macc@nucleus}{}_{}$}%
		\dimen@\wd\tw@
		\advance\dimen@-\wd\z@
		\divide\dimen@ 3
		\@tempdima\wd\tw@
		\advance\@tempdima-\scriptspace
		\divide\@tempdima 10
		\advance\dimen@-\@tempdima
		\ifdim\dimen@>\z@ \dimen@0pt\fi
		\rel@kern{0.6}\kern-\dimen@
		\if#31
		\overline{\rel@kern{-0.6}\kern\dimen@\macc@nucleus\rel@kern{0.4}\kern\dimen@}%
		\advance\dimen@0.4\dimexpr\macc@kerna
		\let\final@kern#2%
		\ifdim\dimen@<\z@ \let\final@kern1\fi
		\if\final@kern1 \kern-\dimen@\fi
		\else
		\overline{\rel@kern{-0.6}\kern\dimen@#1}%
		\fi
	}%
	\macc@depth\@ne
	\let\math@bgroup\@empty \let\math@egroup\macc@set@skewchar
	\mathsurround\z@ \frozen@everymath{\mathgroup\macc@group\relax}%
	\macc@set@skewchar\relax
	\let\mathaccentV\macc@nested@a
	\if#31
	\macc@nested@a\relax111{#1}%
	\else
	\def\gobble@till@marker##1\endmarker{}%
	\futurelet\first@char\gobble@till@marker#1\endmarker
	\ifcat\noexpand\first@char A\else
	\def\first@char{}%
	\fi
	\macc@nested@a\relax111{\first@char}%
	\fi
	\endgroup
}
\makeatother
\makeatletter
\newcommand*{\rom}[1]{\expandafter\@slowromancap\romannumeral #1@}
\makeatother

\makeatletter
\newcommand{\roml}[1]{\lowercase\expandafter{\romannumeral #1\relax}}
\makeatother

\DeclareMathAlphabet{\mathpzc}{OT1}{pzc}{m}{it}

\newcommand{\ket}[1]{| #1 \rangle}
\newcommand{\bra}[1]{\langle #1 |}

\newcommand{\V}{\mathcal{V}}

\begin{document}
\title{Characterizing a high-dimensional unitary transformation without measuring the qudit it transforms}
\author{Salini Rajeev}
	\affiliation{Department of Physics, 145 Physical Sciences Bldg., Oklahoma State University, Stillwater, OK 74078, USA.}
	
	\author{Mayukh Lahiri}
	\email{mlahiri@okstate.edu} \affiliation{Department of Physics, 145 Physical Sciences Bldg., Oklahoma State University, Stillwater, OK 74078, USA.}

% use {asbstract*} to suppress the copyright line. Copyright information will be added in production

\begin{abstract} 
We present a method for reconstructing an arbitrary high-dimensional unitary transformation without detecting the qudit that it transforms. We demonstrate the method using orbital angular momentum states of light. Our method relies on quantum interference enabled by path identity of undetected photons. The method is practically useful when suitable detectors are not available for the qudit on which the unitary transformation works. 
\end{abstract}
%%%%%%%%%%%%%%%%%%%%%%%%%%  body  %%%%%%%%%%%%%%%%%%%%%%%%%%

\maketitle

The evolution of a state in quantum mechanics is often described using a unitary transformation. Measurement of unitary transformations therefore allows us to study the dynamics of a quantum system. Measuring unitary transformations is also crucial for quantum information science because operations of quantum gates are usually described by unitary transformations \cite{nielsen2010quantum}. We consider the problem of measuring high-dimensional unitary transformations that are applied to single qudit states. Since the dimension of the Hilbert space gives the fundamental resource of quantum information processing \cite{blume2002climbing}, qudit states (high-dimensional) are more desirable than qubit states (two-dimensional) for encoding and processing information. Consequently, there have been efforts to characterize high-dimensional unitary transformations (see, for example, \cite{baldwin2014quantum,escandon2024estimation}).
\par
Standard techniques of measuring unitary transformations rely on the principle of quantum process tomography \cite{poyatos1997complete,chuang1997prescription}, in which the measurement is performed on the transformed quantum state. Direct detection of the transformed state is required even when the process is ancilla-assisted \cite{altepeter2003ancilla}. We propose an alternative approach in which we avoid the detection of the transformed qudit. Our method would be advantageous in situations where adequate detectors are not available for the transformed qudit. Such situations often arise with photonic qudits since adequate single-photon detectors are not available for a wide spectral range.
\par
Our method is based on a unique quantum interference phenomenon that was first demonstrated by Zou, Wang, and Mandel \cite{zou1991induced}. This type of interference is also called \emph{interference by path identity} \cite{hochrainer2022quantum} and has found applications in quantum imaging \cite{lemos2014quantum, lahiri2015theory}, spectroscopy \cite{kalashnikov2016infrared}, and sensing \cite{qian2023quantum,dalvit2024quantum}.
\par
\begin{figure}[ht!]
\centering\includegraphics[width=\linewidth]{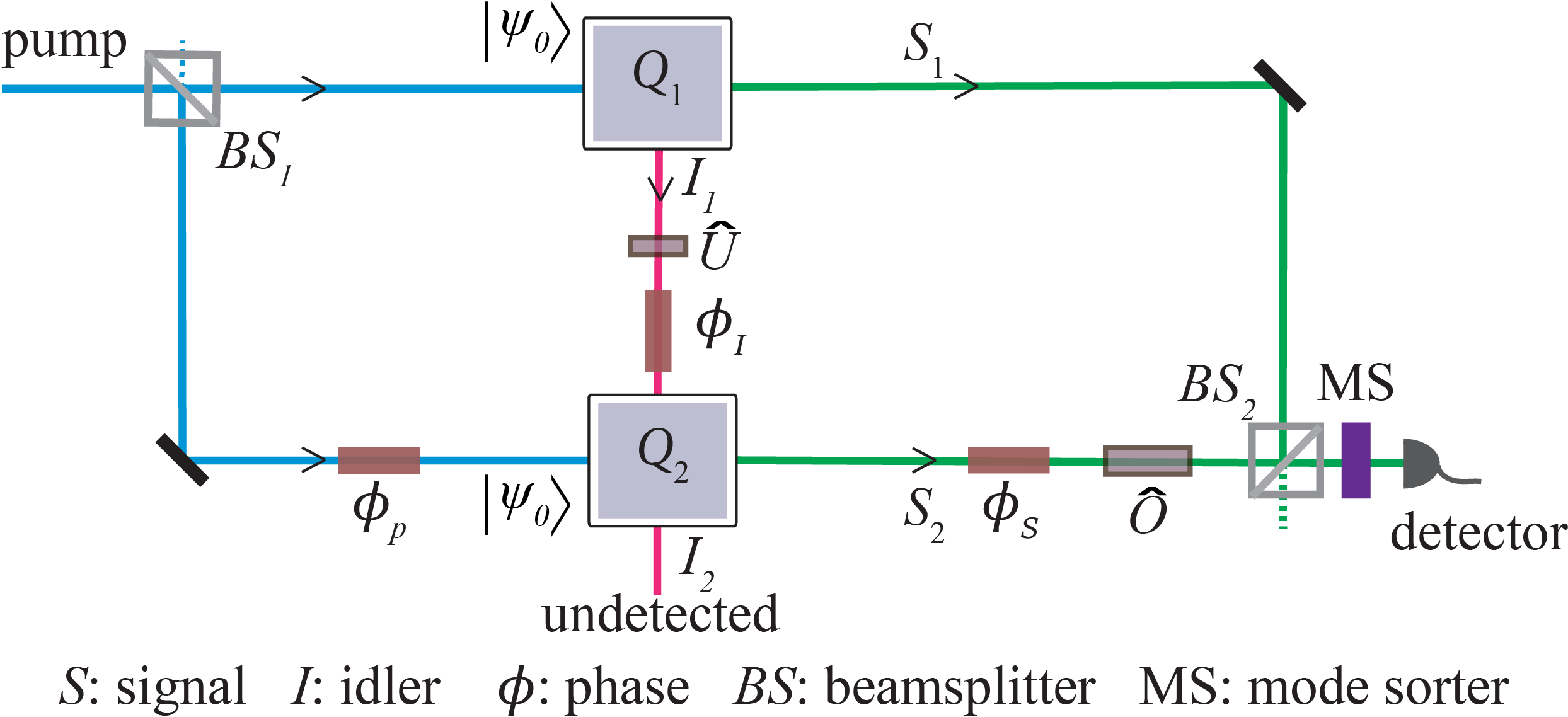}
\caption{High-dimensional unitary transformation measurement scheme. Two sources, $Q_1$ and $Q_2$, emit coherently. Each can emit an OAM-entangled photon pair ($I,S$) in two-qudit state $\ket{\psi_0}$ [Eq.~(\ref{q-state})]. Beams of $S$-photons ($S_1$ and $S_2$) are superposed by a beamsplitter ($BS_2$), and single-photon counting rate is measured at an output after each $S$-photon is projected onto a suitable OAM mode. $I$-photon beam $I_1$ is sent through $Q_2$ and aligned with $I$-photon beam $I_2$. This alignment (path identity) makes beams $S_1$ and $S_2$ mutually coherent and single-photon interference patterns are observed at the detector. An unknown unitary transformation ($\widehat{U}$) is applied to $I$-photon in beam $I_1$. Its information appears in single-photon interference patterns obtained by detecting $S$-photons. $\widehat{U}$ can be fully reconstructed from these interference patterns if suitable unitary transformations ($\widehat{O}$) are applied to $S$-photons in beam $S_2$.} \label{fig:setup}
\end{figure}

We work with the orbital angular momentum (OAM) states of photons. We consider a Zou-Wang-Mandel interferometer that contains two identical sources $Q_1$ and $Q_2$ (Fig.~\ref{fig:setup}). 
Each source can generate a photon pair (signal and idler) in the following state: 
\begin{align}\label{q-state}
    \ket{\psi_0}=\frac{1}{\sqrt{N}}\sum_{l=0}^{N-1}\ket{-l_I,l_S},
\end{align}
where $\ket{-l_I,l_S}\equiv\ket{-l_I}\otimes\ket{l_S}$ with $\ket{-l_I}$ and $\ket{l_S}$ representing the states of idler ($I$) and signal ($S$) photons in OAM modes $-l$ and $l$, respectively. Such states can be generated by spontaneous parametric down conversion (SPDC) nonlinear crystals.
Signal and idler photons can have different frequencies (wavelengths). The signal photon is detected, whereas the idler photon remains undetected.
 \par
Source $Q_j$ (with $j=1,2$) emits the photons $I$ and $S$ into the paths $I_j$ and $S_j$, respectively.
We assume that the probability of multiple pair emission as well as contributions from stimulated emissions are negligible \cite{path-id-note}. In this case, the quantum state generated by the two sources is a linear superposition of the states generated by them individually. The state is thus given by
\begin{align}\label{total-q-state}
    \ket{\Tilde{\psi}}=\frac{1}{\sqrt{2N}}\sum_{l=0}^{N-1}\left(\ket{-l_{I_1},l_{S_1}}+e^{i\phi_p}\ket{-l_{I_2},l_{S_2}}\right),
\end{align}
where $\phi_p$ represents a phase (e.g., the phase difference between the pump fields at the two nonlinear crystals) and we have assumed for simplicity that the sources emit with the same probability.
\par
Beam $I_1$ is sent through the source $Q_2$ and is aligned with the beam $I_2$ as shown in Fig.~\ref{fig:setup}. Such an alignment makes the paths of the idler photon emerging from the two sources identical and is sometimes called \emph{path identity} \cite{hochrainer2022quantum}. On its way from $Q_1$ to $Q_2$, an unknown unitary transformation, $\widehat{U}$, is applied to the idler photon in the $I_1$ beam. \emph{This transformation ($\widehat{U}$) is to be reconstructed.} 
\par
The quantization of OAM modes in the \emph{paraxial limit} \cite{calvo2006quantum,karimi2014radial,plick2013quantum}, allows one to write $\ket{-l_{I_j}}=\widehat{a}^\dag_{-l_{I_j}}\ket{\text{vac}}$, where $\ket{\text{vac}}$ represents the vacuum state and $\widehat{a}^\dag_{-l_{I_j}}$ is the creation operator corresponding to the idler photon in OAM mode $-l$ created by source $Q_j$ $ (j=1,2)$.
The application of the unitary transformation $\widehat{U}$ to the idler field in beam $I_1$ together with the path identity (Fig.~\ref{fig:setup}) results in the relation
    \begin{align}\label{align-cond}
        \widehat{a}_{-l_{I_2}}=e^{i\phi_I}\sum_{l'=0}^{N-1}U_{ll'}\widehat{a}_{-l'_{I_1}},
    \end{align}
where $U_{ll'}$ represents matrix elements of the unitary transformation to be measured, and $\phi_I$ is the phase gained by the idler field while traveling from $Q_1$ to $Q_2$. Upon applying the Hermitian conjugate of Eq.~(\ref{align-cond}) to $\ket{\text{vac}}$, we obtain the following relation:
    \begin{align}\label{align-cond-kets}
        \ket{-l_{I_2}}=e^{-i\phi_I}\sum_{l'=0}^{N-1}U^*_{ll'}\ket{-l'_{I_1}}.
    \end{align}
Using Eqs.~(\ref{align-cond}) and (\ref{align-cond-kets}), we find that the quantum state generated by the two sources with path identity is given by
\begin{align}\label{total-q-state-algn}
    \ket{\psi}=&\frac{1}{\sqrt{2N}}\sum_{l=0}^{N-1} \Big(\ket{-l_{I_1},l_{S_1}}+e^{-i(\phi_I-\phi_p)}\nonumber\\&\times\sum_{l'=0}^{N-1}U^*_{ll'}\ket{-l'_{I_1},l_{S_2}}\Big).
\end{align}
\par
A known unitary transformation, $\widehat{O}$, which can be controlled experimentally, is applied to the signal photon in beam $S_2$. The two signal beams $S_1$ and $S_2$ are then superposed by a balanced beamsplitter ($BS_2$), and one of its outputs is sent to a detector. Each signal photon is projected onto a chosen OAM state before detection. Therefore, the quantum field at the detector can be represented by
\begin{align}\label{e-field}
    \widehat{E}^{(+)}_{l} \propto \widehat{a}_{l_{S_1}}+i \, e^{i\phi_S}\sum_{k=0}^{N-1}O_{lk} \, \widehat{a}_{k_{S_2}},
\end{align}
where $\widehat{a}_{l_{S_j}}$ $(j=1,2)$ is the annihilation operator corresponding to a signal photon in the OAM mode $l$ emitted by source $Q_j$, and $\phi_S$ is the phase difference corresponding to the difference between the optical paths along the beams $S_1$ and $S_2$.
\par
The single-photon counting rate for mode $l$, which is linearly proportional to the probability of detecting a signal photon in the OAM state $\ket{l_S}$, is given by
$P_{l} \propto \bra{\psi}\widehat{E}^{(-)}_{l}\widehat{E}^{(+)}_l\ket{\psi}$.
It now follows from Eqs.~(\ref{total-q-state-algn}) and (\ref{e-field}) that
\begin{align}\label{ph-counting-rate-gen}
    P_l\propto 1+\sum_{k=0}^{N-1}|O_{lk}||U_{kl}|\sin(\phi_{in}+\text{arg}\{U_{kl}\}-\text{arg}\{O_{lk}\}),
\end{align}
where $\phi_{in}=\phi_I-\phi_p-\phi_S$. It is evident from Eq.~(\ref{ph-counting-rate-gen}) that when $\phi_{in}$ is varied, $P_l$ varies sinusoidally. Therefore, $P_l$ represents a single-photon interference pattern. We note that the interference pattern is obtained without any coincidence measurement or postselection. The cause of this interference is that the two signal beams $S_1$ and $S_2$ become mutually coherent due to the path identity of the idler photon. We observe in Eq.~(\ref{ph-counting-rate-gen}) that the complete information of the unknown unitary transformation ($\widehat{U}$) is present in the interference pattern. This happens despite the fact that the idler photon, on which $\widehat{U}$ is applied, is not detected. 
\par
We now show that $\widehat{U}$ can be fully reconstructed from the single-photon interference patterns [Eq.~(\ref{ph-counting-rate-gen})] by suitably choosing $\widehat{O}$. To illustrate the process, we first choose a simple form of $\widehat{O}$, which we call a \emph{basic form} and denote it by $\widehat{O}_b(q,r)$ with $q<r$. We define the matrix representing $\widehat{O}_b(q,r)$ by one that is obtained by replacing four elements ($qq$, $qr$, $rq$, and $rr$) of an $N\times N$ unit matrix with the four elements of a $2\times 2$ rotation matrix in the OAM space. That is, (i)~$[O_{b}(q,r)]_{qq}=[O_{b}(q,r)]_{rr}=\cos\theta$ and $[O_{b}(q,r)]_{rq}=-[O_{b}(q,r)]_{qr}=\sin\theta$, where $\theta$ is an experimentally controllable parameter; and (ii)~$[O_{b}(q,r)]_{ll'}=\delta_{ll'}$ if $l\neq q,r$ and $l'\neq q,r$, where $\delta_{ll'}$ is the Kronecker delta. The matrix form of $\widehat{O}_b(q,r)$ is given by
\begin{align}\label{u-k}
\begin{blockarray}{rccccccc}
 & 0 & \cdots & q & \cdots & r & \cdots & N\!-\!1 \\
\begin{block}{r[ccccccc]}
0     &1 &  \cdots & 0 & \cdots & 0 & \cdots & 0 \\
\vdots & \vdots & \ddots & \vdots &  & \vdots &  & \vdots \\
q   & 0 & \cdots & \underbrace{\cos{\theta}}_{O_{qq}} & \cdots & \underbrace{-\sin{\theta}}_{O_{qr}} & \cdots & 0 \\
\vdots & \vdots & & \vdots & \ddots & \vdots & & \vdots \\
r   & 0 & \cdots & \underbrace{\sin{\theta}}_{O_{rq}} & \cdots & \underbrace{\cos{\theta}}_{O_{rr}} & \cdots & 0 \\
\vdots & \vdots & & \vdots & & \vdots & \ddots & \vdots \\
 N\!-\!1  & 0 & \cdots & 0 & \cdots & 0 & \cdots & 1 \\
\end{block}
\end{blockarray}.  
\end{align}
\par
It is evident from Eq.~(\ref{u-k}) that $\widehat{O}_b(q,r)$ performs a two-dimensional rotational transformation on OAM modes $q$ and $r$ of the signal photon while leaving the remaining modes unchanged. Let us now project the signal photons emerging from $BS_2$ (Fig.~\ref{fig:setup}) onto OAM modes $q$ and $r$. In this case, using Eqs.~(\ref{ph-counting-rate-gen}) and (\ref{u-k}) we find that 
\begin{subequations}
    \begin{align}
        P_q^{(q,r)}(\theta)&\propto 1+\cos \theta |U_{qq}|\sin[\phi_{in} +\text{arg}\{U_{qq}\}]+\sin \theta \nonumber\\&+|U_{rq}|\sin[\phi_{in}-\pi+\text{arg}\{U_{rq}\}],\label{p-h-q}\\
        P_r^{(q,r)}(\theta)&\propto 1+\cos \theta |U_{rr}|\sin[\phi_{in}+\text{arg}\{U_{rr}\}]+\sin \theta\nonumber\\&+|U_{qr}|\sin[\phi_{in}+\text{arg}\{U_{qr}\}],\label{p-h-r}
    \end{align}
\end{subequations}
where the superscript $(q,r)$ signifies that the signal OAM modes $q$ and $r$ have undergone rotational transformation. 
From Eqs.~(\ref{p-h-q}) and (\ref{p-h-r}), we obtain the following set of interference patterns: 
\begin{subequations}
    \begin{align}
        &P_{q}^{(q,r)}(\theta=0) \propto 1 + |U_{qq}|\sin[\phi_{in} +\text{arg}\{U_{qq}\}],\label{p-h-q-1} \\
        &P_{r}^{(q,r)}(\theta=0) \propto 1 +|U_{rr}|\sin[\phi_{in}+\text{arg}\{U_{rr}\}] \label{p-h-r-1}, \\
        &P_{q}^{(q,r)}(\theta=\pi/2) \propto 1 +|U_{rq}|\sin[\phi_{in}-\pi+\text{arg}\{U_{rq}\}], \label{p-h-q-2} \\
        &P_{r}^{(q,r)}(\theta=\pi/2) \propto 1 +|U_{qr}|\sin[\phi_{in}+\text{arg}\{U_{qr}\}] \label{p-h-r-2},
    \end{align}
\end{subequations}
Equations (\ref{p-h-q-1})--(\ref{p-h-r-2}) show that these single-photon interference patterns contain complete information (both amplitude and argument) of two diagonal elements ($U_{qq},U_{rr}$) and two off-diagonal elements ($U_{qr},U_{rq}$) of the unknown unitary transformation. These matrix elements can be readily determined using standard interferometric procedure. The amplitudes of these matrix elements are given by the visibility \cite{vis-note} of these interference patterns [Eqs.~(\ref{p-h-q-1})-(\ref{p-h-r-2})], i.e., by 
\begin{subequations}
    \begin{align}
        &\V_{q}^{(q,r)}(\theta=0)=|U_{qq}|, \hspace{0.5 cm} \V_{q}^{(q,r)}(\theta=\pi/2)=|U_{rq}|,\label{v-q}\\
        &\V_{r}^{(q,r)}(\theta=0)=|U_{rr}|, \hspace{0.5 cm} \V_{r}^{(q,r)}(\theta=\pi/2)=|U_{qr}|,\label{v-r}
    \end{align}
\end{subequations}
where $\V$ denotes visibility. The arguments of these matrix elements can, for example, be determined from the phase differences between these interference patterns (for an illustration see Fig.~\ref{phase-plots-illustration} below).
\par
It is worth noting that if the signal photons emerging from $BS_2$ are projected onto OAM modes $l\neq q,r$ before detection, Eqs.~(\ref{ph-counting-rate-gen}) and (\ref{u-k}) yield the following set of interference patterns:
\begin{equation}\label{p-h-l}
P_{l\neq q,r} \propto 1 +|U_{ll}|\sin(\phi_{in}+\text{arg}\{U_{ll}\}). 
\end{equation}
Therefore, the diagonal matrix elements $U_{ll}$ (with $l\neq q,r$) can also be determined in addition to the four matrix elements ($U_{qq},U_{rr}$,$U_{qr},U_{rq}$) for a single choice of $\widehat{O}_b(q,r)$.
\par
\begin{figure}[htbp]
\centering\includegraphics[width=\linewidth]{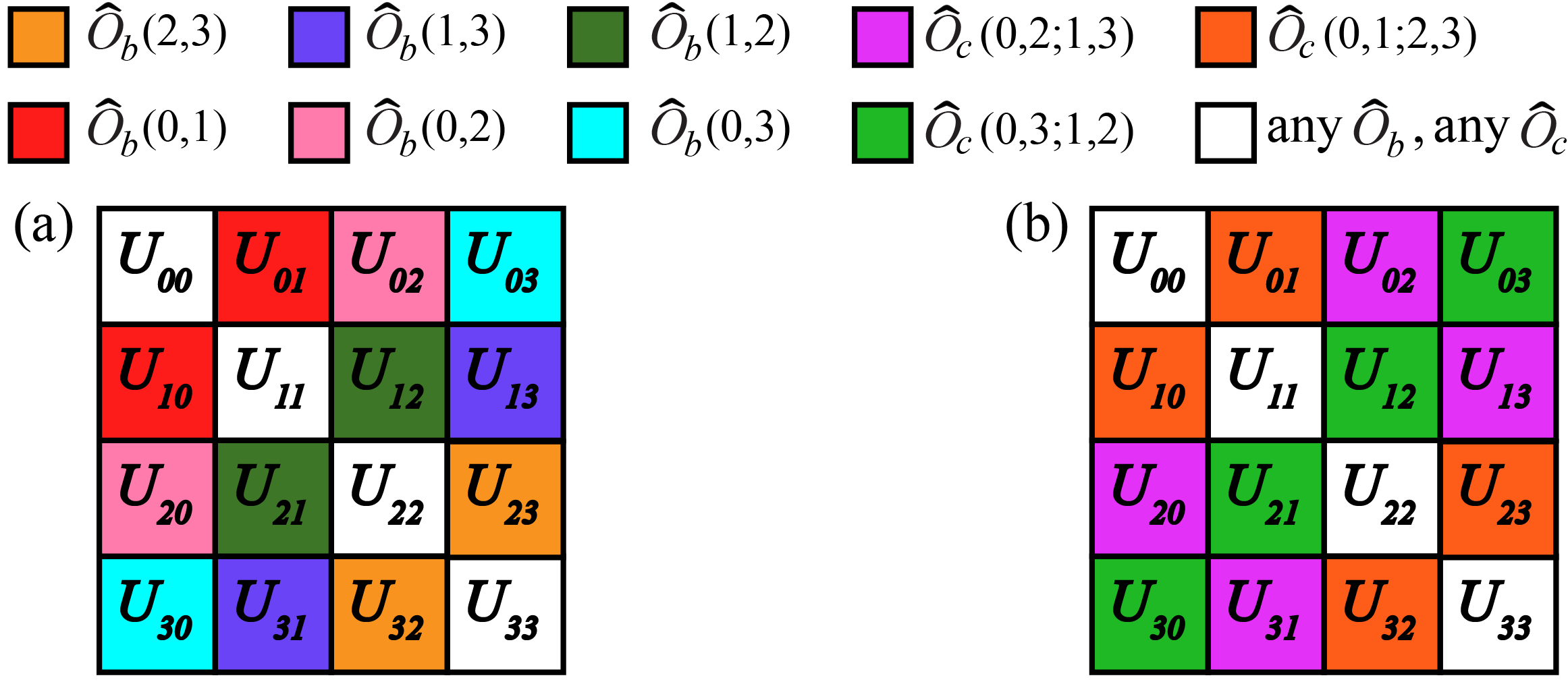}
\caption{Basic and compound forms of the known unitary transformation ($\widehat{O}$) required to measure a four-dimensional unitary transformation ($\widehat{U}$). (a) Each cell represents a matrix element of $\widehat{U}$. The color of the cell identifies the basic form of $\widehat{O}$ required to determine the matrix element represented by the cell. (b) The color of a cell shows the compound form of $\widehat{O}$ required to determine the matrix element represented by the cell.}
\label{four-dimensional-illustration}
\end{figure}

The results shown in Eqs.~(\ref{p-h-q-1})--(\ref{p-h-r-2}) and Eq.~(\ref{p-h-l}) allow one to develop several protocols to determine all matrix elements of the unknown unitary transformation. The conceptually simplest protocol can be constructed using Eqs.~(\ref{p-h-q-1})--(\ref{p-h-r-2}) only. One can first choose $q=0$; since $q<r$, the corresponding possible values of $r$ are given by $r=1,2,\dots, N-1$. Therefore, for $q=0$, there are $N-1$ basic forms of $\widehat{O}$, which are $\widehat{O}_b(0,1), \widehat{O}_b(0,2), \dots, \widehat{O}_b(0,N-1)$. In each case, $4$ matrix elements of $\widehat{U}$ are determined using Eqs.~(\ref{p-h-q-1})--(\ref{p-h-r-2}). However, the element $U_{00}$ is obtained in each case. Therefore, using these $N-1$ choices of $\widehat{O}_b(q,r)$, one can determine $4(N-1)-(N-2)=3N-2$ matrix elements of $\widehat{U}$; note that these matrix elements contain all diagonal elements. One can now increase the value of $q$ by steps of $1$ and in each step, follow the same procedure until all matrix elements of $\widehat{U}$ are obtained. It can be checked that in order to determine all the matrix elements of $\widehat{U}$, one needs to consider $N(N-1)/2$ basic forms of $\widehat{O}$. For example, in the four-dimensional case, one needs $4\times 3/2=6$ choices, which are $\widehat{O}_b(0,1)$, $\widehat{O}_b(0,2)$, $\widehat{O}_b(0,3)$, $\widehat{O}_b(1,2)$, $\widehat{O}_b(1,3)$, and $\widehat{O}_b(2,3)$ (Fig.~\ref{four-dimensional-illustration} (a)).
\par
We note that more than one basic form of $\widehat{O}$ [i.e., $\widehat{O}_b(q,r)$] can be combined by multiplication into a single unitary transformation. We call the resulting transformation a \emph{compound form} of $\widehat{O}$. However, while combining a number of $\widehat{O}_b(q,r)$ in this manner, it must be ensured that a specific OAM mode of the signal photon is acted upon by only one basic form of $\widehat{O}$. For example, we multiply two basic forms, $\widehat{O}_b(q_1,r_1)$ and $\widehat{O}_b(q_2,r_2)$, only when $q_1\neq q_2,r_2$ and $r_1\neq q_2,r_2$. The resulting compound form in this case is represented by $\widehat{O}_c(q_1,r_1;q_2,r_2)=\widehat{O}_b(q_1,r_1)\widehat{O}_b(q_2,r_2)$; note that this multiplication is commutative. Such compound forms of $\widehat{O}$ can equivalently be used to reconstruct the matrix elements of $\widehat{U}$. To illustrate it, we now consider a four-dimensional system.
\par
As mentioned above, for a four-dimensional system, there are $6$ basic forms of $\widehat{O}$, which can be used separately to determine all matrix elements of $\widehat{U}$ (Fig.~\ref{four-dimensional-illustration}a). Let us now multiply two basic forms, $\widehat{O}_b(0,1)$ and $\widehat{O}_b(2,3)$, to construct a compound form $\widehat{O}_c(0,1;2,3)=\widehat{O}_b(0,1)\widehat{O}_b(2,3)$. The matrix form of $\widehat{O}_c(0,1;2,3)$ is given by (see Appendix A)
\begin{align}\label{u-k-4:a}
\widehat{O}_c(0,1;2,3)
=\begin{pmatrix}
\cos\theta & -\sin\theta & 0 & 0 \\
\sin\theta & \cos\theta & 0 & 0 \\
0 & 0 & \cos\theta & -\sin\theta \\
0 & 0 & \sin\theta & \cos\theta 
\end{pmatrix}.
\end{align}
Substituting for the matrix elements of $\widehat{O}$ from Eq.~(\ref{u-k-4:a}) into Eq.~(\ref{ph-counting-rate-gen}), we obtain the following eight single-photon interference patterns: $P_0^{(0,1)}(\theta=0)$, $P_1^{(0,1)}(\theta=0)$, $P_0^{(0,1)}(\theta=\pi/2)$, $P_1^{(0,1)}(\theta=\pi/2)$, $P_2^{(2,3)}(\theta=0)$, $P_3^{(2,3)}(\theta=0)$, $P_2^{(2,3)}(\theta=\pi/2)$, and $P_3^{(2,3)}(\theta=\pi/2)$. Using these interference patterns, we can determine the following eight matrix elements of the unknown unitary transformation: $U_{00}$, $U_{11}$, $U_{10}$, $U_{01}$, $U_{22}$, $U_{33}$, $U_{32}$, and $U_{23}$, respectively (Appendix A). As expected, these elements are also obtained by using the basic forms $\widehat{O}_b(0,1)$ and $\widehat{O}_b(2,3)$ separately. The remaining matrix elements of $\widehat{U}$ can be likewise determined by using two more compound forms of $\widehat{O}$: $\widehat{O}_c(0,2;1,3)=\widehat{O}_b(0,2)\widehat{O}_b(1,3)$ and $\widehat{O}_c(0,3;1,2)=\widehat{O}_b(0,3)\widehat{O}_b(1,2)$ (Appendix A).
That is, instead of six basic forms of $\widehat{O}$ (Fig.~\ref{four-dimensional-illustration}a), we can equivalently use three compound forms of $\widehat{O}$ to construct all matrix elements of an arbitrary four-dimensional unitary transformation (Fig.~\ref{four-dimensional-illustration}b). 
\par
To numerically illustrate our results, we now reconstruct the matrix elements of the four-dimensional Hadamard gate \cite{brandt2020high}, which has the matrix form
\begin{align}\label{Hadamard}
    \widehat{U}=\frac{1}{2}\begin{pmatrix}
    1&1&1&1\\
    1&i&-1&-i\\
    1&-1&1&-1\\
    1&-i&-1&i
        \end{pmatrix}.
\end{align}
We obtain $16$ relevant interference patterns [Appendix B, Eqs.~(\ref{p-h-a-0-1-had})-(\ref{p-h-c-3-2-had})] using the three above-mentioned compound forms of $\widehat{O}$: $\widehat{O}_c(0,1;2,3)$, $\widehat{O}_c(0,2;1,3)$, and $\widehat{O}_c(0,3;1,2)$. We find that the visibility \cite{vis-note} of each of these interference patterns is $1/2$, which implies that $|U_{lm}|=1/2, ~\forall ~l,m$ (three examples are illustrated in Fig.~\ref{phase-plots-illustration}). The arguments of all these matrix elements are also determined using these interference patterns. 
\begin{figure}[ht!]
\centering\includegraphics[width=\linewidth]{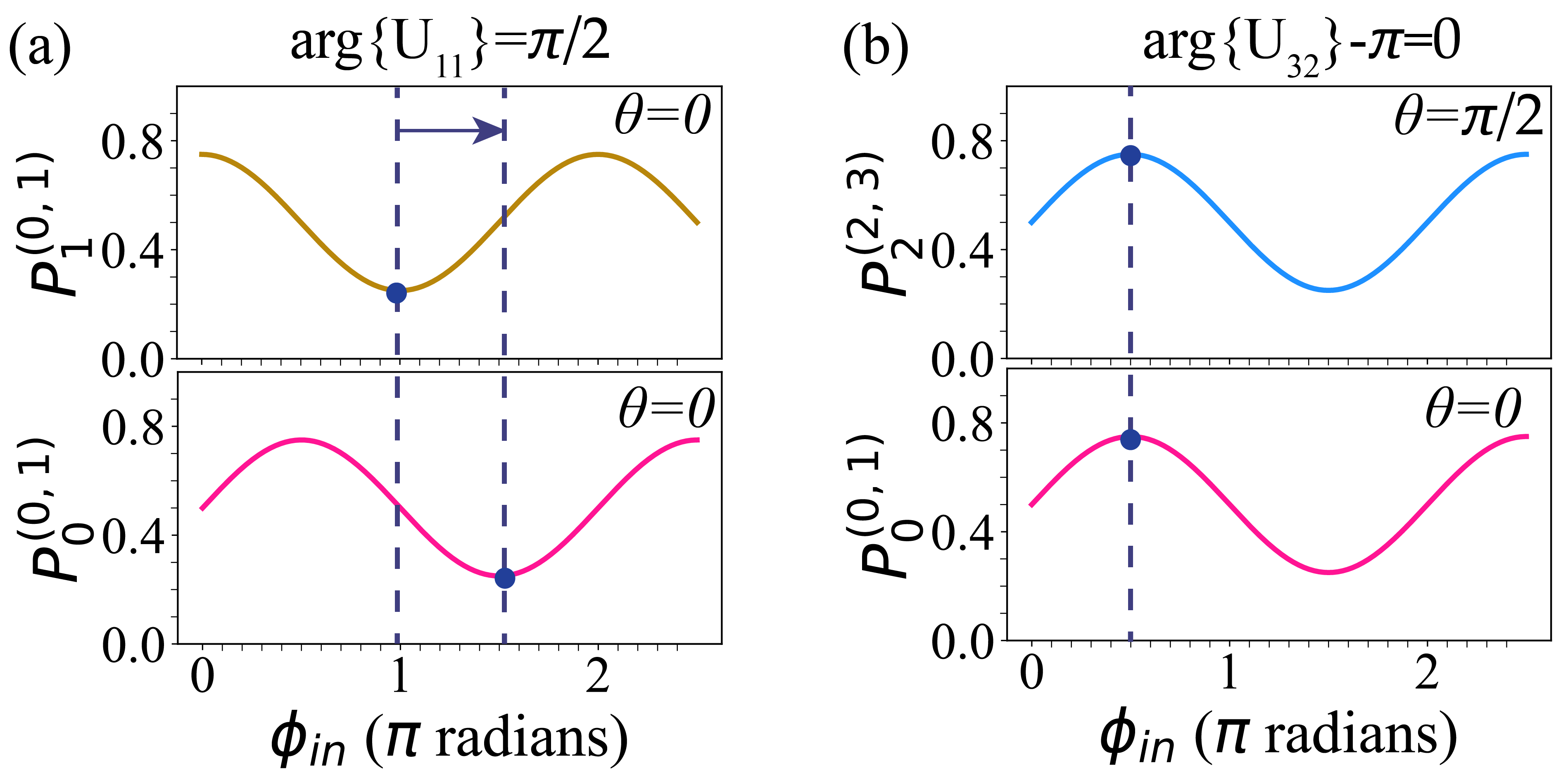}
\caption{Illustrating determination of matrix elements of a four-dimensional Hadamard gate. \textbf{Arguments:} (a) The phase difference between the interference patterns $P_1^{(0,1)}(\theta=0)$ and $P_0^{(0,1)}(\theta=0)$ gives the $\text{arg}\{U_{11}\}$, which is equal to $\pi/2$. (b) The phase difference between the interference patterns $P_2^{(2,3)}(\theta=\pi/2)$ and $P_0^{(0,1)}(\theta=0)$ is zero, implying $\text{arg}\{U_{32}\}-\pi=0$,  i.e., $\text{arg}\{U_{32}\}=\pi$. \textbf{Modulus:} Visibilities of $P_0^{(0,1)}(\theta=0)$, $P_1^{(0,1)}(\theta=0)$, and $P_2^{(2,3)}(\theta=\pi/2)$ are all equal to $1/2$, implying that $|U_{00}|=1/2$, $|U_{11}|=1/2$, and $|U_{32}|=1/2$, respectively.}
\label{phase-plots-illustration}
\end{figure}
As examples, we illustrate retrieval of phases $\text{arg}\{ U_{11} \}$ and $\text{arg}\{ U_{32} \}$ in Fig.~\ref{phase-plots-illustration}. When we compare the interference patterns $P_1^{(0,1)}(\theta=0)$ and $P_0^{(0,1)}(\theta=0)$ (Appendix B, Eqs.~(\ref{p-h-a-0-1-had}) and (\ref{p-h-a-1-1-had})), we find that the phase difference between them is $\pi/2$ (Fig.~\ref{phase-plots-illustration}a), which is equal to $\text{arg}\{U_{11}\}$. Similarly, for the four-dimensional Hadamard gate, the phase difference between the interference patterns $P_2^{(2,3)}(\theta=\pi/2)$ and $P_0^{(0,1)}(\theta=0)$ (Appendix B, Eqs.~(\ref{p-h-a-0-1-had}) and (\ref{p-h-a-2-2-had})) turns out to be zero, which must be equal to $\text{arg}\{U_{32}\}-\pi$. Consequently, we have $\text{arg}\{U_{32}\}=\pi$, which matches with the value given in Eq.~(\ref{Hadamard}). In the same manner, arguments of all matrix elements of the Hadamard gate are retrieved (Appendix B). 
\par
We now briefly discuss a prescription for combining multiple basic forms of $\widehat{O}$ for a general $N$-dimensional system. As already mentioned, there are $N(N-1)/2$ basic forms of $\widehat{O}$ in this case. While combining them, we need to ensure that a specific signal OAM-mode is not acted upon by more than one such basic form. Therefore, the maximum number of basic forms of $\widehat{O}$ that can be combined by multiplication to form a compound form is $m=\left\lfloor \frac{N}{2} \right\rfloor$, where $\left\lfloor\hspace{0.15 cm} \right\rfloor$ denotes floor division \cite{floor-div-note}. An example of such a compound form is $\widehat{O}_c(q_1,r_1; \dots ;q_m,r_m)=\widehat{O}_b(q_1,r_1)\widehat{O}_b(q_2,r_2)\cdots\widehat{O}_b(q_m,r_m)$, where $q_j\neq q_k, r_k$ and $r_j\neq q_k, r_k$ with $j\neq k$ for all $j,k=1,\dots, m$. Using this compound form of $\widehat{O}$, we can determine all diagonal elements and the following off-diagonal elements of the unknown unitary transformation: $U_{q_1r_1}$, $U_{r_1q_1}$, $U_{q_2r_2}$, $U_{r_2q_2}$, $\cdots$, $U_{q_mr_m}$, $U_{r_mq_m}$. Reconstructing all matrix elements of $\widehat{U}$ requires $N-1$ such compound forms for even $N$, and $N$ such forms for odd $N$. When $N=4$, the maximum number of basic forms that can be combined is $m=\left\lfloor \frac{4}{2} \right\rfloor=2$ and the required number of compound forms is $N-1=3$, as we have found above. 
\par
In summary, our results show that it is possible to measure any high-dimensional unitary transformation without detecting the photon that it transforms. Recent experimental implementations of high-dimensional unitary transformations for OAM modes of light \cite{schlederer2016cyclic,brandt2020high} strongly suggest that our method is experimentally implementable. Since our analysis is based on quantum field theory, our method can be extended to non-photonic quantum systems.

\section*{Acknowledgment} The research was supported by the Air Force Office of Scientific Research under grant FA9550-23-1-0216.
\appendix
\section{Measurement of four-dimensional unitary transformations using compound forms of $\widehat{O}$}
We begin by deriving Eq.~(13) in the main text, which represents the matrix form of $\widehat{O}_c(0,1;2,3)$. This compound form of $\widehat{O}$ is obtained by multiplying the basic forms $\widehat{O}_b(0,1)$ and $\widehat{O}_b(2,3)$, i.e.,
\begin{widetext}
\begin{align}
    \underbrace{\begin{pmatrix}
    \cos\theta & -\sin\theta&0&0\\
    \sin\theta & \cos\theta &0&0\\
    0&0&1&0\\
    0&0&0&1 
    \end{pmatrix}}_{\widehat{O}_b(0,1)}\cdot \underbrace{\begin{pmatrix}
    1&0&0&0\\
    0&1&0&0\\
    0&0&\cos\theta & -\sin\theta\\
    0&0&\sin\theta & \cos\theta
    \end{pmatrix}}_{\widehat{O}_b(2,3)}=\underbrace{\begin{pmatrix}
    \cos\theta & -\sin\theta & 0 & 0 \\
    \sin\theta & \cos\theta & 0 & 0 \\
    0 & 0 & \cos\theta & -\sin\theta \\
    0 & 0 & \sin\theta & \cos\theta 
\end{pmatrix}}_{\widehat{O}_c(0,1;2,3)}.\label{comp-form-1}
\end{align}
\end{widetext}
We now show that $\widehat{O}_c(0,1;2,3)$ can be used to determine eight matrix elements of the unknown unitary transformation $\widehat{U}$.
Using Eqs.~(\ref{comp-form-1}) and (7) of the main text, and choosing $\theta$ appropriately, we obtain the following eight single-photon interference patterns :
\begin{subequations}
    \begin{align}
        &P_{0}^{(0,1)} (\theta=0)\propto 1+|U_{00}|\sin(\phi_{in}+\text{arg}\{U_{00}\}),\label{p-h-a-0-1}\\
        &P_{0}^{(0,1)} (\theta=\pi/2)\propto 1-|U_{10}|\sin(\phi_{in}+\text{arg}\{U_{10}\}),\label{p-h-a-0-2}\\
        &P_{1}^{(0,1)}(\theta=0)\propto 1+|U_{11}|\sin(\phi_{in}+\text{arg}\{U_{11}\})\label{p-h-a-1-1},\\
        &P_{1}^{(0,1)}(\theta=\pi/2)\propto 1+|U_{01}|\sin(\phi_{in}+\text{arg}\{U_{01}\})\label{p-h-a-1-2},\\
        &P_{2}^{(2,3)}(\theta=0)\propto 1+|U_{22}|\sin(\phi_{in}+\text{arg}\{U_{22}\})\label{p-h-a-2-1},\\
        &P_{2}^{(2,3)}(\theta=\pi/2)\propto 1-|U_{32}|\sin(\phi_{in}+\text{arg}\{U_{32}\})\big]\label{p-h-a-2-2},\\
        &P_{3}^{(2,3)}(\theta=0)\propto 1+|U_{33}|\sin(\phi_{in}+\text{arg}\{U_{33}\})\label{p-h-a-3-1},\\
        &P_{3}^{(2,3)}(\theta=\pi/2)\propto 1+|U_{23}|\sin(\phi_{in}+\text{arg}\{U_{23}\})\label{p-h-a-3-2}.
    \end{align}
\end{subequations}
Equations (\ref{p-h-a-0-1})-(\ref{p-h-a-3-2}) show that the single-photon interference patterns contain the complete information of the eight matrix elements $U_{00}$, $U_{10}$, $U_{11}$, $U_{01}$, $U_{22}$, $U_{32}$, $U_{33}$, and $U_{23}$. These matrix elements can be readily determined using the standard interferometric procedure as discussed in the main text.
\par
We now construct the remaining two compound forms of $\widehat{O}$, which are $\widehat{O}_c(0,2;1,3)=\widehat{O}_b(0,2)\widehat{O}_b(1,3)$ and $\widehat{O}_c(0,3;1,2)=\widehat{O}_b(0,3)\widehat{O}_b(1,2)$.  Their matrix forms are obtained as follows:
\begin{widetext}
\begin{subequations}
    \begin{align}
    &\underbrace{\begin{pmatrix}
    \cos\theta &0& -\sin\theta&0\\
    0&1&0&0\\
    \sin\theta &0& \cos\theta &0\\
    0&0&0&1 
    \end{pmatrix}}_{\widehat{O}_b(0,2)}\cdot \underbrace{\begin{pmatrix}
    1&0&0&0\\
    0&\cos\theta &0& -\sin\theta\\
    0&0&1&0\\
    0&\sin\theta&0 & \cos\theta
    \end{pmatrix}}_{\widehat{O}_b(1,3)}=\underbrace{\begin{pmatrix}
    \cos\theta &0& -\sin\theta&0\\
    0&\cos\theta & 0&-\sin\theta\\
    \sin\theta &0& \cos\theta &0\\
    0&\sin\theta & 0&\cos\theta 
    \end{pmatrix}}_{\widehat{O}_c(0,2;1,3)},\label{comp-form-2}\allowdisplaybreaks\\
    &\underbrace{\begin{pmatrix}
    \cos\theta &0&0& -\sin\theta\\
    0&1&0&0\\
    0&0&1&0\\
    \sin\theta &0&0& \cos\theta\\
    \end{pmatrix}}_{\widehat{O}_b(0,3)}\cdot \underbrace{\begin{pmatrix}
    1&0&0&0\\
    0&\cos\theta & -\sin\theta&0\\
    0&\sin\theta& \cos\theta&0\\
        0&0&0&1
    \end{pmatrix}}_{\widehat{O}_b(1,2)}=\underbrace{\begin{pmatrix}
    \cos\theta &0&0& -\sin\theta\\
    0&\cos\theta &-\sin\theta&0\\
    0&\sin\theta &\cos\theta &0\\
    \sin\theta &0& 0&\cos\theta 
    \end{pmatrix}}_{\widehat{O}_c(0,3;1,2)}\label{comp-form-3}.
    \end{align}
\end{subequations} 
\end{widetext}
Using Eqs.~(\ref{comp-form-2}), (\ref{comp-form-3}) and (7) of the main text, and choosing $\theta$ appropriately, we obtain the following relevant equations:
\begin{subequations}
    \begin{align}
     &\text{for~} \widehat{O}_c(0,2;1,3):\nonumber\\
        &P_{0}^{(0,2)} (\theta=\pi/2)\propto 1-|U_{20}|\sin(\phi_{in}+\text{arg}\{U_{20}\}),\label{p-h-b-0-2}\\
        &P_{1}^{(1,3)}(\theta=\pi/2)\propto 1-|U_{31}|\sin(\phi_{in}+\text{arg}\{U_{31}\})\label{p-h-b-1-2},\\
        &P_{2}^{(0,2)}(\theta=\pi/2)\propto 1+|U_{02}|\sin(\phi_{in}+\text{arg}\{U_{02}\})\big]\label{p-h-b-2-2},\\
        &P_{3}^{(1,3)}(\theta=\pi/2)\propto 1+|U_{13}|\sin(\phi_{in}+\text{arg}\{U_{13}\})\label{p-h-b-3-2},\allowdisplaybreaks\\
    &\text{for~} \widehat{O}_c(0,3;1,2):\nonumber\\
        &P_{0}^{(0,3)} (\theta=\pi/2)\propto 1-|U_{30}|\sin(\phi_{in}+\text{arg}\{U_{30}\}),\label{p-h-c-0-2}\\
        &P_{1}^{(1,2)}(\theta=\pi/2)\propto 1-|U_{21}|\sin(\phi_{in}+\text{arg}\{U_{21}\})\label{p-h-c-1-2},\\
        &P_{2}^{(1,2)}(\theta=\pi/2)\propto 1+|U_{12}|\sin(\phi_{in}+\text{arg}\{U_{12}\})\big]\label{p-h-c-2-2},\\
        &P_{3}^{(0,3)}(\theta=\pi/2)\propto 1+|U_{03}|\sin(\phi_{in}+\text{arg}\{U_{03}\})\label{p-h-c-3-2}.
    \end{align}
\end{subequations}
We observe that the interference patterns represented by Eqs.~(\ref{p-h-b-0-2})-(\ref{p-h-c-3-2}) contain both amplitudes and arguments of the remaining eight matrix elements: $U_{20}$, $U_{31}$, $U_{02}$, $U_{13}$, $U_{30}$, $U_{21}$, $U_{12}$, and $U_{03}$. 

\section{Calculations for the reconstruction of the matrix form of a four-dimensional Hadamard gate}
A four dimensional Hadamard gate has the matrix form:
\begin{align}
    \widehat{U}=\frac{1}{2}\begin{pmatrix}
    1&1&1&1\\
    1&i&-1&-i\\
    1&-1&1&-1\\
    1&-i&-1&i
        \end{pmatrix}.
\end{align}
We use Eqs.~(\ref{p-h-a-0-1})-(\ref{p-h-a-3-2}) and (\ref{p-h-b-0-2})-(\ref{p-h-c-3-2}) to obtain expressions for the relevant single-photon interference patterns. These interference patterns are given by (the proportionality constant is chosen $1/2$)
\begin{subequations}
    \begin{align}
       &\text{for~} \widehat{O}_c(0,1;2,3):\nonumber\\
        &P_{0}^{(0,1)} (\theta=0)=\frac{1}{2}\left[1+\frac{1}{2}\sin(\phi_{in})\right],\label{p-h-a-0-1-had}\\
        &P_{0}^{(0,1)} (\theta=\pi/2)=\frac{1}{2}\left[1-\frac{1}{2}\sin(\phi_{in})\right],\label{p-h-a-0-2-had}\\
        &P_{1}^{(0,1)} (\theta=0)=\frac{1}{2}\left[1+\frac{1}{2}\sin(\phi_{in}+\pi/2)\right]\label{p-h-a-1-1-had},\\
        &P_{1}^{(0,1)} (\theta=\pi/2)=\frac{1}{2}\left[1+\frac{1}{2}\sin(\phi_{in})\right]\label{p-h-a-1-2-had},\\
        &P_{2}^{(2,3)}(\theta=0)=\frac{1}{2}\left[1+\frac{1}{2}\sin(\phi_{in})\right]\label{p-h-a-2-1-had},\\
        &P_{2}^{(2,3)}(\theta=\pi/2)=\frac{1}{2}\left[1-\frac{1}{2}\sin(\phi_{in}+\pi)\right]\label{p-h-a-2-2-had},\\
        &P_{3}^{(2,3)}(\theta=0)=\frac{1}{2}\left[1+\frac{1}{2}\sin(\phi_{in}+\pi/2)\right]\label{p-h-a-3-1-had},\\
        &P_{3}^{(2,3)}(\theta=\pi/2)=\frac{1}{2}\left[1+\frac{1}{2}\sin(\phi_{in}+\pi)\right]\label{p-h-a-3-2-had},\\
        &\text{for~} \widehat{O}_c(0,2;1,3):\nonumber\\
        &P_{0}^{(0,2)} (\theta=\pi/2)=\frac{1}{2}\left[1-\frac{1}{2}\sin(\phi_{in})\right],\label{p-h-b-0-2-had}\\
        &P_{1}^{(1,3)} (\theta=\pi/2)=\frac{1}{2}\left[1-\frac{1}{2}\sin(\phi_{in}-\pi/2)\right]\label{p-h-b-1-2-had},\\
        &P_{2}^{(0,2)}(\theta=\pi/2)=\frac{1}{2}\left[1+\frac{1}{2}\sin(\phi_{in})\right]\label{p-h-b-2-2-had},\\
        &P_{3}^{(1,3)}(\theta=\pi/2)=\frac{1}{2}\left[1+\frac{1}{2}\sin(\phi_{in}-\pi/2)\right]\label{p-h-b-3-2-had},\\
        &\text{for~} \widehat{O}_c(0,3;1,2):\nonumber\\
        &P_{0}^{(0,3)} (\theta=\pi/2)=\frac{1}{2}\left[1-\frac{1}{2}\sin(\phi_{in})\right],\label{p-h-c-0-2-had}\\
        &P_{1}^{(1,2)} (\theta=\pi/2)=\frac{1}{2}\left[1-\frac{1}{2}\sin(\phi_{in}+\pi)\right]\label{p-h-c-1-2-had},\\
        &P_{2}^{(1,2)}(\theta=\pi/2)=\frac{1}{2}\left[1+\frac{1}{2}\sin(\phi_{in}+\pi)\right]\label{p-h-c-2-2-had},\\
        &P_{3}^{(0,3)}(\theta=\pi/2)=\frac{1}{2}\left[1+\frac{1}{2}\sin(\phi_{in})\right]\label{p-h-c-3-2-had}.
    \end{align}
\end{subequations}
\begin{table*}
\setlength{\tabcolsep}{5 pt} 
\renewcommand{\arraystretch}{1} 
\begin{tabular}{|c|cc|cc|cc|}
\hline 
visibility&\multicolumn{2}{c|}{$\widehat{O}_c(0,1;2,3)$}& \multicolumn{2}{c|}{$\widehat{O}_c(0,2;1,3)$} &  \multicolumn{2}{c|}{$\widehat{O}_c(0,3;1,2)$}  \\
\hline
& $\theta=0$& $\theta=\pi/2$& $\theta=0$& $\theta=\pi/2$& $\theta=0$& $\theta=\pi/2$\\
\hline
$\V_{0}$&$|U_{00}|=\frac{1}{2}$&$|U_{10}|=\frac{1}{2}$&$|U_{00}|=\frac{1}{2}$&$|U_{20}|=\frac{1}{2}$&$|U_{00}|=\frac{1}{2}$&$|U_{30}|=\frac{1}{2}$\\
$\V_{1}$&$|U_{11}|=\frac{1}{2}$&$|U_{01}|=\frac{1}{2}$&$|U_{11}|=\frac{1}{2}$&$|U_{31}|=\frac{1}{2}$&$|U_{11}|=\frac{1}{2}$&$|U_{21}|=\frac{1}{2}$\\
$\V_{2}$&$|U_{22}|=\frac{1}{2}$&$|U_{32}|=\frac{1}{2}$&$|U_{22}|=\frac{1}{2}$&$|U_{02}|=\frac{1}{2}$&$|U_{22}|=\frac{1}{2}$&$|U_{12}|=\frac{1}{2}$\\
$\V_{3}$&$|U_{33}|=\frac{1}{2}$&$|U_{23}|=\frac{1}{2}$&$|U_{33}|=\frac{1}{2}$&$|U_{13}|=\frac{1}{2}$&$|U_{33}|=\frac{1}{2}$&$|U_{03}|=\frac{1}{2}$\\
\hline 
\end{tabular}
\caption{Values of $|U_{lm}|$ obtained from the single-photon interference patterns using the three different compound forms of $\widehat{O}$. Visibility of a single-photon interference pattern for the OAM mode $l$ is denoted by $\V_l$.}
\label{tab-illustration-mag}
\end{table*}    
\begin{table*}
\setlength{\tabcolsep}{5 pt} 
\renewcommand{\arraystretch}{1}
\begin{tabular}{cccc}
\hline
arguments & compound form & Interference patterns & value\\
\hline
$\text{arg}\{U_{10}\}$ & $\widehat{O}_c(0,1;2,3)$ & $P_0^{(0,1)}(\theta=\pi/2)$, $P_0^{(0,1)}(\theta=0)$ & 0\\
$\text{arg}\{U_{11}\}$ & $\widehat{O}_c(0,1;2,3)$ & $P_1^{(0,1)}(\theta=0)$, $P_0^{(0,1)}(\theta=0)$ & $\pi/2$\\
$\text{arg}\{U_{01}\}$ & $\widehat{O}_c(0,1;2,3)$ & $P_1^{(0,1)}(\theta=\pi/2)$, $P_0^{(0,1)}(\theta=0)$ & 0\\
$\text{arg}\{U_{22}\}$ & $\widehat{O}_c(0,1;2,3)$ & $P_2^{(2,3)}(\theta=0)$, $P_0^{(0,1)}(\theta=0)$ & $0$\\
$\text{arg}\{U_{32}\}$ & $\widehat{O}_c(0,1;2,3)$ & $P_2^{(2,3)}(\theta=\pi/2)$, $P_0^{(0,1)}(\theta=0)$ & $\pi$\\
$\text{arg}\{U_{33}\}$ & $\widehat{O}_c(0,1;2,3)$ & $P_3^{(2,3)}(\theta=0)$, $P_0^{(0,1)}(\theta=0)$ & $\pi/2$\\
$\text{arg}\{U_{23}\}$ & $\widehat{O}_c(0,1;2,3)$ & $P_3^{(2,3)}(\theta=\pi/2)$, $P_0^{(0,1)}(\theta=0)$ & $\pi$\\
$\text{arg}\{U_{20}\}$ & $\widehat{O}_c(0,2;1,3)$ & $P_0^{(0,2)}(\theta=\pi/2)$, $P_0^{(0,2)}(\theta=0)$ & 0\\
$\text{arg}\{U_{02}\}$ & $\widehat{O}_c(0,2;1,3)$ & $P_2^{(0,2)}(\theta=\pi/2)$, $P_0^{(0,2)}(\theta=0)$ & 0\\
$\text{arg}\{U_{31}\}$ & $\widehat{O}_c(0,2;1,3)$ & $P_1^{(1,3)}(\theta=\pi/2)$, $P_0^{(0,2)}(\theta=0)$ & $-\pi/2$\\
$\text{arg}\{U_{13}\}$ & $\widehat{O}_c(0,2;1,3)$ & $P_3^{(1,3)}(\theta=\pi/2)$, $P_0^{(0,2)}(\theta=0)$ & $-\pi/2$\\
$\text{arg}\{U_{30}\}$ & $\widehat{O}_c(0,3;1,2)$ & $P_0^{(0,3)}(\theta=\pi/2)$, $P_0^{(0,3)}(\theta=0)$ & 0\\
$\text{arg}\{U_{03}\}$ & $\widehat{O}_c(0,3;1,2)$ & $P_3^{(0,3)}(\theta=\pi/2)$, $P_0^{(0,3)}(\theta=0)$ & 0\\
$\text{arg}\{U_{21}\}$ & $\widehat{O}_c(0,3;1,2)$ & $P_1^{(1,2)}(\theta=\pi/2)$, $P_0^{(0,3)}(\theta=0)$ & $\pi$\\
$\text{arg}\{U_{12}\}$ & $\widehat{O}_c(0,3;1,2)$ & $P_2^{(1,2)}(\theta=\pi/2)$, $P_0^{(0,3)}(\theta=0)$ & $\pi$\\
\hline
\end{tabular}
\caption{Values of $\text{arg}\{U_{lm}\}$ obtained by comparing different single-photon interference patterns using the three different compound forms of $\widehat{O}$.}
\label{tab-illustration-phase}
\end{table*}    
The magnitudes of matrix elements are obtained from the visibility of the corresponding interference patterns. In Table.~\ref{tab-illustration-mag}, we show the magnitudes of all retrieved matrix elements of the four-dimensional Hadamard gate.
\par
The arguments of all matrix elements are obtained by comparing relevant pairs of single-photon interference patterns as discussed and demonstrated by examples in the main text (see Fig.~3 in the main text). In Table.~\ref{tab-illustration-phase}, we display the arguments of the retrieved matrix elements of the four-dimensional Hadamard gate.

%\clearpage
% Bibliography
\bibliography{ref}

\begin{thebibliography}{22}
\bibitem[{vis()}]{vis-note}
\bibinfo{note}{The visibility is determined by the standard formula
  $\mathcal{V}=([P]_{\text{max}}-[P]_{\text{min}})/([P]_{\text{max}}+[P]_{\text{min}})$.}
\bibitem[{flo()}]{floor-div-note}
\bibinfo{note}{Floor division is a division operation performed between two
  integers returning the largest integer less than or equal to the quotient.}
\end{thebibliography}

% Full bibliography added automatically for Optics Letters submissions; the following line will simply be ignored if submitting to other journals.
% Note that this extra page will not count against page length
%\bibliographyfullrefs{ref}

\end{document}